\documentclass[prl,aps,twocolumn,showpacs,preprintnumbers,amsmath,amssymb]{revtex4}

\usepackage{graphicx}
\usepackage{dcolumn}
\usepackage{bm}

\begin{document}

\preprint{APS/123-QED}

\title{Dynamics of F=2 Spinor Bose-Einstein Condensates}
\author{H.~Schmaljohann}
\author{M.~Erhard}
\author{J.~Kronj\"ager}
\author{M.~Kottke$^1$}
\author{S.~van~Staa}
\author{J.~J.~Arlt$^1$}
\author{K.~Bongs}
\author{K.~Sengstock}

\affiliation{Institut f\"ur Laser-Physik, Universit\"at Hamburg,
Luruper Chaussee 149, 22761 Hamburg, Germany\\$^1$ Institut f\"ur
Quantenoptik, Universit\"at Hannover, Welfengarten 1, 30167 Hannover,
Germany}

\date{\today}

\begin{abstract}
We experimentally investigate and analyze the rich dynamics in F=2 spinor
Bose-Einstein condensates of $^{87}$Rb. An interplay
between mean-field driven spin dynamics and hyperfine-changing
losses in addition to interactions with the thermal component is observed. In
particular we measure conversion rates in the range of 
$10^{-12}\,\mbox{cm}^3\mbox{s}^{-1}$ for spin changing collisions within
the F=2 manifold and spin-dependent loss rates in the range of
$10^{-13}\, \mbox{cm}^3\mbox{s}^{-1}$ for hyperfine-changing
collisions. From our data we observe a polar behavior in the F=2 ground state of
$^{87}$Rb, while we measure the F=1 ground state to be ferromagnetic.
Furthermore we see a magnetization for condensates prepared with
non-zero total spin.
\end{abstract}

\pacs{03.75.Mn, 34.50.Pi, 03.75.Hh}


\maketitle The investigation of atomic spin systems is central for the
understanding of magnetism and a highly active area of research
e.g.~with respect to magnetic nanosystems, spintronics and
magnetic interactions in high T$_c$ superconductors. In addition
entangled spin systems in atomic quantum gases show intriguing
prospects for quantum optics and quantum
computation~\cite{Pu2000,You2000a,Soerensen2001,Julsgaard2001a,Mandel2003a}.
Bose-Einstein condensates (BEC) of ultra-cold atoms offer new regimes
for studies of collective spin
phenomena~\cite{Myatt1997a,Hall1998a,Matthews1998a,
McGuirk2002a,Stenger1999a,Miesner1999a,Stamper-Kurn1999b,Leanhardt2003}.
BECs with spin degree of freedom are special 
in the sense that their order parameter is a vector in contrast 
to the ``common'' BEC where it is a scalar. Recent extensive studies 
have been made in optically trapped $^{23}$Na in the F=1
state~\cite{Stenger1999a, Miesner1999a,Stamper-Kurn1999b,Leanhardt2003}. In
addition evidence of spin dynamics was demonstrated in optically
trapped $^{87}$Rb in the F=1 state~\cite{Barrett2001a}. 
There is current interest in extending the systems under investigation to
F=2 spinor condensates~\cite{Goerlitz2003a,Ciobanu2000a,Ho2000a,
Koashi2000,Klausen2001a,Ueda2002a}, which add significant new
physics. F=2 spinor condensates offer richer dynamics, an
additional magnetic phase, the so-called cyclic phase~\cite{Koashi2000,Ciobanu2000a}, as well as intrinsic
connections to d-wave superconductors~\cite{Mermin1974}.

In this letter we present first studies of optically trapped 
$^{87}$Rb F=2 spinor condensates. We measure rates for spin changing
collisions for different channels within the F=2 manifold and
discuss the steady state for various initial
conditions. Additionally we observe and discuss
the thermalization of dynamically populated $m_F$ condensates. 
We also present measurements of spin-dependent hyperfine 
decay rates of the F=2 state in $^{87}$Rb, as a key to further 
understanding the intensively studied collisional properties of 
$^{87}$Rb~\cite{Kempen2002,Marte2002}.

\begin{figure}[h]
\centering
\includegraphics[width=1.0\linewidth]{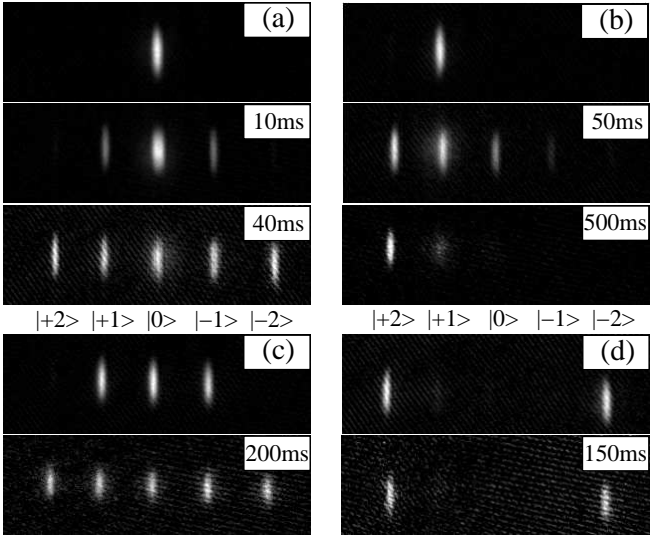}
\caption{Time-dependent observation of different $m_F$ components starting from
the initially prepared states denoted by (a)-(d).
 Shown are spinor condensates separated by a Stern-Gerlach method (time of flight 31ms).\label{f:pictures}}
\end{figure}

Our experimental setup consists of a compact double MOT apparatus
which produces magnetically trapped $^{87}$Rb
Bose-Einstein condensates containing $10^6$ atoms in the F=2,
$m_F=2$ state. To confine the atoms independently
of their spin state they are subsequently transferred into a far 
detuned optical dipole trap. It is operated at 1064$\, $nm generating trapping 
frequencies of typically $2\pi\times891\, $Hz vertically,
$2\pi\times155\, $Hz horizontally and $2\pi\times21.1\, $Hz along the beam
direction. After transfer we further cool the ensemble for $500\,
$ms by selective parametric excitation~\cite{Poli2002} resulting
in approximately $10^5$ optically trapped atoms and a condensate
fraction well above 60\%. We are able to prepare arbitrary spin
compositions using rapid adiabatic passage and controlled Landau-Zener 
crossing techniques~\cite{Mewes1997a} at an offset field
around $25\, $G. 
After initial state preparation the magnetic field is lowered to a
value of typically $340(\pm 20) \,$mG, with a field gradient below
$15\, $mG/cm, to ensure a well-defined quantization axis and a good 
overlap of the different $m_F$ states during spin dynamics~\cite{Note}. 
Spin dynamics is subsequently allowed to proceed during a variable
hold time: Stored in the optical trap the condensate spin degree of
freedom can evolve under well-controlled conditions. 
Due to interatomic interactions, initially prepared $m_F$ components 
can evolve into other $m_F$ components, e.g.~by processes like 
$|0\rangle+|0\rangle\leftrightarrow|\!+\!1\rangle+|\!-\!1\rangle$, 
while - disregarding atomic losses - the total spin is conserved. 
This process is expected to end in the spinor ground state distribution
showing the hyperfine dependent magnetic properties of the atomic
species under investigation. Experimentally we detect different
spin components by spatially separating them with a Stern-Gerlach
method during time of flight (after switching off the trapping
potential).
 Absorption imaging is then used to evaluate the respective spatial density distributions 
as well as the number of atoms in the condensate and thermal fractions for
each spin component. Figure~\ref{f:pictures} shows typical spinor condensate
evolutions for different starting conditions.
These pictures demonstrate a rich spectrum of spin dynamics 
in F=2 systems, caused by an intriguing interplay
of spin component coupling and spin-dependent losses. In the
following we will investigate these processes separately.

We analyze the observed spinor evolution following a mean-field
approach~\cite{Ho1998a,Koashi2000,Ciobanu2000a}, in which the
properties of a spinor condensate are determined by a
spin-dependent energy functional. Extending the approach of
~\cite{Koashi2000} by the experimentally relevant quadratic Zeeman
energy term we determine the following functional for F=2 systems:
\begin{equation} \label{e:energy}
K_{spin}=c_1\langle \mbox{\boldmath$F$} \rangle^2+\frac{4}{5}
c_2|\langle s_- \rangle|^2-p\langle F_z\rangle-q\langle
F_z^2\rangle ,
\end{equation}
where $\langle \mbox{\boldmath$F$} \rangle $, $\langle F_z\rangle
$ and $\langle s_- \rangle $ denote the expectation values for the
spin vector, its $z$ component and the spin-singlet pair amplitude.
The experimentally controllable parameters $p$ and $q$ span the
phase space for the system. Here $p$ represents the mean spin
of the system while $q$ is the quadratic Zeeman energy,
representing the magnetic offset field.

The spin-dependent mean-field is characterized by the parameters
$c_1=\frac{4\pi\hbar^2}{m}\cdot\frac{a_4-a_2}{7}$ and
$c_2=\frac{4\pi\hbar^2}{m}\cdot\frac{7a_0-10a_2+3a_4}{7}$,
introducing the s-wave scattering length $a_f$ for collisions
involving a total spin $f$ of the colliding pair. In the following
the $F=2,m_F=X$ state is denoted by $|X\rangle$. For simplicity relative 
phases in mixtures are neglected. 
The $c_1$ term includes all couplings of states with $\Delta m_F=\pm1$, e.g.~
$|0\rangle+|0\rangle\leftrightarrow |\!+\!1\rangle+|\!-\!1\rangle$. 
The $c_2$ term includes the only possible coupling with $\Delta m_F=\pm2$~: 
$|0\rangle+|0\rangle\leftrightarrow|\!+\!2\rangle+|\!-\!2\rangle$. 
The magnitude of these terms is connected to the timescales of spin dynamics 
and their relative strengths indicate the initially dominant channels. Taking the
calculated scattering lengths and our initial experimental
conditions (offset field $B=340\, $mG, mean density
$n\approx1.1\times 10^{14}\,\mbox{cm}^{-3}$) the 
energy ranges for the different terms are: spin independent mean-field 
$k_B\times64\, $nK, spin-dependent mean-field $c_1:k_B\times0..12\,$nK, 
$c_2:k_B\times0..0.05\,$nK and quadratic Zeeman energy $q:k_B\times0..1.6\,$nK.

Minimizing the energy functional, Eq.~(\ref{e:energy}) leads to the 
parameter-dependent ground state spin composition of the
system, reflecting its magnetic properties~\cite{wirselbst}. Theory predicts
$^{87}$Rb in F=2 to be in the polar phase, but close to the border to
the so-called cyclic phase~\cite{Klausen2001a}.

\begin{figure}
\centering
\includegraphics[width=1.0\linewidth]{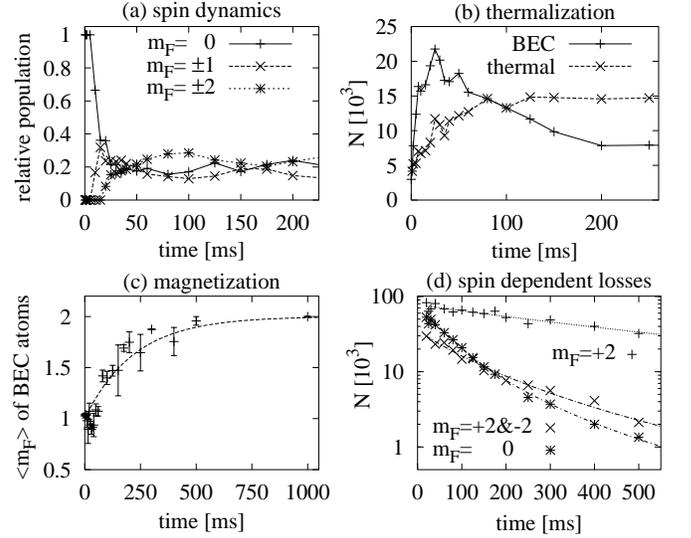}
\caption{Spin dynamics and spin-dependent losses. (a) Relative population of
the $m_F$ states of a condensate initially prepared in $|0\rangle$. 
(b) Thermalization of a BEC in the $|\!+\!2\rangle$ state form initially $|0\rangle+|\!+\!1\rangle$. 
(c) Total spin of the condensate initially prepared in $|\!+\!1\rangle$. 
(d) Decay curves for total number of condensed atoms in all $m_F$ states, given for different inital $m_F$ preparations.
\label{f:graphs}}
\end{figure}

Experimentally we find the system dominated by three processes,
setting different timescales, in increasing order: spin dynamics,
two-body hyperfine losses and three-body recombination. This
hierarchy allows us to separately analyze individual interaction
processes, which are exemplarily given in Fig.~\ref{f:graphs}.

In a first set of measurements we study the evolution of
spinor condensates starting from different initial spin
states. Figure~\ref{f:graphs}(a) shows the evolution of the relative
occupation of the different $m_F$ components of a F=2 spinor
condensate initially prepared in the state $|0\rangle $, also
shown in Fig.~\ref{f:pictures}(a). This example
clearly demonstrates that spin evolution takes place predominantly
between neighboring spin states, as expected from coupling via the
dominating $c_1$-term discussed above. Thus first the
components $|\!\pm\! 1\rangle$ and only afterwards the
$|\!\pm\! 2\rangle$ states are populated. Interestingly there is a delay in the
occupation of $|\!\pm\! 1\rangle$ and a further delay in the
occupation of $|\!\pm\! 2\rangle$ states. The latter delay can be
intuitively explained since there must first be some occupation in
the $|\!\pm\! 1\rangle$ states before the fully stretched states can
be occupied. The origin of the initial delay as well as the
''overshoot'' of the $|\!\pm\! 1\rangle$ occupation is still under
investigation.

As an interesting feature Fig.~\ref{f:graphs}(a) shows
evidence of spinor oscillations on a 100$\, $ms timescale around 
equipartition. This indicates a coherent evolution between the 
different $m_F$ components, with only minor influence of the thermal
cloud on this timescale.

In fact we always observe that spin dynamics is significantly quicker than
thermalization. Spin dynamics thus leads to the formation of pure
condensate wavefunctions in new spin states, which subsequently
thermalize on a much longer timescale. This opens a new way to
study thermalization effects with an independently tunable heat
bath. In Fig.~\ref{f:graphs}(b) the condensate and thermal fractions of the newly created 
$|\!+\!2\rangle$ state, starting with a mixture of the
states $|\!+\!1\rangle$ and $|0\rangle$ are plotted. At first the condensate 
in the $|\!+\!2\rangle$ state is almost pure. This can be intuitively
understood, as spin dynamics predominantly occurs in the high
density condensate fraction on a timescale of the order of 25$\,
$ms. The subsequent buildup of a thermal cloud in the newly
populated spin states takes place within a significantly longer time,
of the order of 100$\, $ms. Dominant effects contributing
to the thermalization of the new spin state are the interaction of
the new spin condensate with the parent thermal cloud
(''melting''), and spin dynamics of the parent condensate fraction
with the parent thermal cloud leading to a direct occupation of
the new component thermal cloud. In this respect spinor condensates are a possible 
new tool for investigations of finite temperature effects.

In order to draw a more complete picture of spin dynamics in F=2
systems we now investigate the evolution of a variety of initial
states as summarized in Table~\ref{t:rates}. In this table we give
estimates for the initial spin dynamics rates for different
channels as well as the experimentally observed ''final''
distribution.
\begin{table}
\caption{Spin evolution for different preparations.\label{t:rates}}
\begin{tabular}{{cccc}}
\hline \hline
\bf initially&\bf initial &\bf initial channels    &\bf finally\\
\bf prepared&\bf total&\bf into $m_F$ state  &\bf populated\\
\bf $m_F$ states&\bf spin&\bf G $[10^{-13} \mbox{cm}^3 \mbox{s}^{-1}]$ &\bf $m_F$ states\\ \hline 
$|0\rangle$ &0& $\rightarrow |\!\pm\!1\rangle\approx 21.0$ & equipartition\\
$|\!+\!1\rangle+|\!-\!1\rangle$ &0&$\rightarrow |0\rangle\approx 26.9$ & equipartition\\
  &      & $\rightarrow |\!\pm\!2\rangle \approx 4.6$ &\\
$|\!+\!1\rangle+|0\rangle+|\!-\!1\rangle$&0& $\rightarrow |\!\pm\!2\rangle \approx 5.0$ & equipartition\\
$|\!+\!2\rangle+|\!-\!2\rangle$  &0& - & $|\!+\!2\rangle+|\!-\!2\rangle$\\
$|\!+\!2\rangle+|0\rangle+|\!-\!2\rangle$ &0& $\rightarrow |\!\pm\!1\rangle < 0.1$  & $|\!+\!2\rangle+|\!-\!2\rangle$\\
$|\!+\!2\rangle+|\!-\!1\rangle$ &$1/2$&  - & $|\!+\!2\rangle$\\
$|\!+\!1\rangle+|0\rangle$ &$1/2$& $\rightarrow |\!+\!2\rangle\approx 21.7$&$|\!+\!2\rangle$\\
              &     & $\rightarrow |\!-\!1\rangle \approx 19.2$ &\\
$|\!+\!1\rangle$ &1& $\rightarrow |\!+\!2\rangle \approx 22.4$&$|\!+\!2\rangle$\\
      &   & $\rightarrow |0\rangle \approx 12.2$&\\
      &   &($\rightarrow |\!-\!1\rangle \approx 4.7$)&\\
$|\!+\!2\rangle$  &2& - & $|\!+\!2\rangle$\\ \hline \hline
\end{tabular}
\end{table}
The rates are intended for comparison with loss 
rates and between the different spin channel rates. They were obtained
using the total ensemble density, number of BEC atoms in the prepared 
$m_F$ components and the initial slope of a phenomenological exponential 
fit to the respective spin component population curves:
\begin{center}
$G_{\rightarrow|X\rangle}=\frac {dN_{|X\rangle}}{dt}\cdot\frac{N_{initial}}{\langle n_{initial} \rangle}$.
\end{center}
At short timescales we find the rate for $|0\rangle \rightarrow |\!\pm\! 1\rangle$ 
similar to the rate $|\!\pm\! 1\rangle \rightarrow |0\rangle$ as one would expect
for a reversible or coherent process. 
In contrast, at longer timescales the system 
evolves as expected into a time-independent distribution of $m_F$ states. 
As a main result our measurements show the stability of a mixture of 
$|\!\pm\!2\rangle$ (see Table~\ref{t:rates} and Fig.~\ref{f:pictures}(d)), 
for which we also find no spatial separation of the two components in the
trapping volume. This is clear evidence of {\it polar behavior} as
predicted for F=2 spinor condensates of $^{87}$Rb.

Some initial preparations with zero total spin end up with all $m_F$
components equally populated (see also Fig.~\ref{f:pictures}(a,c)). This is in fact 
one of several degenerate ground states of the polar phase 
in the case of zero magnetic field~\cite{Ueda2002a}, but the quadratic Zeeman energy 
lifts this degeneracy. For our experimental parameters a mixture of all $m_F$ components
is not a ground state for any phase in the mean-field description. 
The observed behavior might be due to a lack of time to reach the 
ground state, as spin-dependent losses depopulate the condensate before thermal
equilibrium is reached.

Remarkably, if the atoms are prepared in superpositions which are 
ground states for the {\it cyclic phase} we observe very slow dynamics. 
First the mixture of $|\!+\!2\rangle + |0\rangle + |\!-\!2\rangle$, 
which is the ground state of the cyclic phase with total spin zero~\cite{Ueda2002a} shows 
nearly no spin dynamics. If starting with $|\!+\!2\rangle + |\!-\!1\rangle$ 
as a ground state for the cyclic phase at $B\!>\!0$~\cite{wirselbst} with non-zero spin we do not
 observe any spin dynamics. These cases are particularly surprising,
as the states $|0\rangle$ alone as well as $|\!+\!1\rangle$ alone show fast spin dynamics
as can be seen in Fig.~\ref{f:pictures}(a,b). In the mixtures however, we observe 
only a faster decay of the $|0\rangle$ and $|\!-\!1\rangle$ component 
compared to $|\!\pm\!2\rangle$. Concluding this, the stability of the $|\!\pm\!2\rangle$ mixture 
demontrates polar behavior of F=2 $^{87}$Rb atoms, and in addition the slow dynamics of 
prepared cyclic ground states shows the F=2 state to be close to the cyclic phase~\cite{Note2}. 

An important point in the dynamics of spinor condensates is total spin 
conservation. This is directly observed in the rates for all spin preparations (see rates 
Table~\ref{t:rates}) except in the case starting with $|\!+\!1\rangle$. 
Here we observe a much higher rate for the production of the $|\!+\!2\rangle$ 
atoms than for the ones in $|0\rangle$. This can be understood as follows: 
Imagine the $|0\rangle$ atoms favor to change their spin into $|\!-\!1\rangle$ 
while interacting with an atom in the initial spin state which then changes 
into $|\!+\!2\rangle$. Then the correct rates are determind by 
the measured values $G_{\rightarrow|\!+\!2\rangle}-G_{\rightarrow|\!-\!1\rangle}$ and $G_{\rightarrow|0\rangle}+G_{\rightarrow|\!-\!1\rangle}$.
These two values are $17.7\times10^{-13} \mbox{cm}^3 \mbox{s}^{-1}$ and $16.9\times10^{-13}\mbox{cm}^3 \mbox{s}^{-1}$, 
respectively, and agree within the expected precision.

Investigating spinor condensates with initial spin $\ne 0$ one
finds that the combination of spin dynamics and spin-dependent
hyperfine losses (see below) leads to a loss induced magnetization
during the evolution to the final state (Fig.~\ref{f:pictures}(b),\ref{f:graphs}(c)). 
The remaining condensate always ends up in the fully stretched state if the initial state
was prepared with non-zero total spin or in any other non-symmetric
mixture. This is due to the fact that only the fully stretched component is
immune to hyperfine-changing collisions.
The magnetization process is only inhibited for a symmetric initial
state having total spin zero.

\enlargethispage{1mm}
Next we focus on hyperfine-changing collisions (i.e.~
involving transitions F=2$\rightarrow$F=1) as a special case
of spin relaxation dynamics.
 The release of hyperfine energy in these collisions
leads to immediate loss of the collision partners from the trap.
This loss process usually takes place on very short timescales,
prohibiting the observation of spin dynamics in the upper
hyperfine level, e.g.~of $^{23}$Na~\cite{Goerlitz2003a}. The relatively 
low hyperfine loss rates we find for $^{87}$Rb are due to a 
coincidental destructive interference of decay paths~\cite{Julienne1997a,Myatt1997a}. 
Fig.~\ref{f:graphs}(d) shows the decay of Bose-Einstein condensates 
of $^{87}$Rb prepared in different $m_F$ states/mixtures. 
The decay of the stretched state $|\!+\!2\rangle $
is dominated by three-body recombination and was measured by~\cite{Soding1999a} 
to be $L=1.8\times10^{-29}\mbox{cm}^6\mbox{s}^{-1}$, a value which also fits to our data. 
The equal superposition of the $|\!+\!2\rangle$
and $|\!-\!2\rangle $ states is special in the sense that we
observe no spin dynamics in this case. The hyperfine-changing
collisions can thus only occur via collisions of the type
$|2,+2\rangle + |2,-2\rangle \rightarrow |1,m_1\rangle +
|F,m_2\rangle (F=1,2) $. We deduce the two-body rate $G=6.6
(\pm 0.9) \times 10^{-14} \mbox{cm}^3 \mbox{s}^{-1}$, which leads to a much faster
decay than three-body losses for the pure $|\!+\!2\rangle$
 state (Fig.~\ref{f:graphs}(d)). 
The dependence of the hyperfine-changing collision rate on the
number of decay channels involved becomes obvious observing the
decay of a condensate initially prepared in the state
$|0\rangle$ (see Fig.~\ref{f:graphs}(d)). It is important
to note that in this case spin dynamics more rapidly ($\le 25\,$ms)
distributes the population almost equally over all $m_F$ states. 
The decay curve (of initially $|0\rangle$) thus effectively represents the loss of a spin state equipartition,
which has access to all possible channels. We determine the mean two-body 
decay rate for this case to be $G=10.2(\pm 1.3) \times 10^{-14} \mbox{cm}^3 \mbox{s}^{-1}$.

Finally we have also studied spinor condensates in the F=1
state where we observe slow spin dynamics on a timescale of seconds 
like in the case of $^{23}$Na in F=1~\cite{Stenger1999a}. Starting in the superposition 
state $|1,\pm1\rangle$ we observe the creation of the $|1,0\rangle$ 
component. The final state with all $m_F$ components populated and spacially 
mixed is reached after about seven seconds. According to the phase diagrams of~\cite{Stenger1999a} 
we have thus shown that {\it $^{\it 87}$Rb in the F=1 state 
is ferromagnetic} as predicted by~\cite{Klausen2001a}.

In conclusion we have presented studies of various
aspects of F=2 spinor condensates in $^{87}$Rb. In particular we
have discussed rates for spin dynamics as well as the evolution into the
magnetic ground state. Furthermore we have investigated spin-dependent
hyperfine-changing collisions and thermalization effects
in newly created spin components during spinor evolution. These
studies have demonstrated the feasibility of $^{87}$Rb condensates
as a model for F=2 spin systems, added data to the collisional 
properties of $^{87}$Rb and opened a new path for finite
temperature studies. As a key result we measure $^{87}$Rb in the F=1 
state to be ferromagnetic and observe polar behavior for the F=2 state. 

We acknowledge stimulating contributions by W.~Ertmer and  support from 
the {\it Deutsche Forschungsgemeinschaft} in the SFB 407 and the SPP 1116.


\begin{thebibliography}{31}
\bibitem[{Not({\natexlab{a}})}]{Note}
\bibinfo{note}{Note that the desired initial state is preserved up to this time
  as spin dynamics at high magnetic field values is either energetically
  suppressed or releases enough quadratic Zeeman energy to eject the
  corresponding atoms from the trap.}
\bibitem[{wir()}]{wirselbst}
\bibinfo{note}{Detailed calculations on ground state properties including the
  quadratic Zeeman energy will be presented elsewhere.}
\end{thebibliography}
\end{document}